\let\rECGBKernelLabel\label
\documentclass[aps,prd,preprint,amsmath,amssymb,nofootinbib,eqsecnum,superscriptaddress]{revtex4-2}
\makeatletter
\IfFormatAtLeastTF{2023-06-01}{%
  \let\label\rECGBKernelLabel
  \let\ltx@label\label
}{}
\makeatother
\usepackage{etoolbox}
\makeatletter
\patchcmd{\@bibdataout@aps}{author="08"}{author="48"}{}{%
  \PackageError{recgb}{Unable to configure APS bibliography author format}{}%
}
\patchcmd{\@bibdataout@aps}{author="08"}{author="48"}{}{%
  \PackageError{recgb}{Unable to configure APS short-bibliography author format}{}%
}
\makeatother
\usepackage[colorlinks=true,linkcolor=blue,citecolor=blue,urlcolor=blue]{hyperref}
\usepackage[dvipsnames]{xcolor}
\usepackage{graphicx}
\usepackage{bm}
\usepackage{booktabs}
\usepackage{needspace}
\usepackage[section]{placeins}
\allowdisplaybreaks

\newcommand{\dd}{\mathrm d}
\newcommand{\lap}{\nabla^2}
\newcommand{\ph}{\dot\phi_0}
\newcommand{\php}{\ddot\phi_0}
\newcommand{\Ch}{\bigl(C^{(3)}+2a^2u\psi\bigr)}
\newcommand{\Vn}[1]{V^{(#1)}}
\newcommand{\hS}{\widehat S}
\newcommand{\hV}[1]{\widehat V^{(#1)}}
\newcommand{\TOne}{T^{(1)}}
\newcommand{\TTwo}{T^{(2)}}
\newcommand{\intk}{\int\!\frac{\dd^3k}{(2\pi)^3}}

\newcommand{\fz}[1]{f_{#1}^{(0)}}

\begin{document}

\title{Alleviating cosmological strong coupling with torsion in regularized 4D Einstein--Gauss--Bonnet gravity}

\author{Jianhui Qiu}
\email{jhqiu@ucas.ac.cn}
\affiliation{School of Fundamental Physics and Mathematical Sciences,
Hangzhou Institute for Advanced Study, UCAS, Hangzhou 310024, China}
\author{Ling-Wei Luo}
\email{lwluo.phys@gmail.com}
\affiliation{School of Fundamental Physics and Mathematical Sciences,
Hangzhou Institute for Advanced Study, UCAS, Hangzhou 310024, China}
\author{Chunhui Liu}
\email{liuchunhui22@mails.ucas.ac.cn}
\affiliation{School of Fundamental Physics and Mathematical Sciences,
Hangzhou Institute for Advanced Study, UCAS, Hangzhou 310024, China}
\affiliation{University of Chinese Academy of Sciences, Beijing 100190, China}
\affiliation{Institute of Theoretical Physics, Chinese Academy of Sciences, Beijing 100190, China}
\author{Chao-Qiang Geng}
\email{cqgeng@ucas.ac.cn}
\affiliation{School of Fundamental Physics and Mathematical Sciences,
Hangzhou Institute for Advanced Study, UCAS, Hangzhou 310024, China}
\affiliation{Department of Physics and Synergetic Innovation Center for Quantum Effects and Applications,
Hunan Normal University, Changsha, Hunan 410081, China}

\begin{abstract}
The four-dimensional scalar--tensor theory obtained by conformally regularizing the Gauss--Bonnet term has a vanishing kinetic term for the scalar perturbation mode on its late-time de Sitter attractor, signaling strong coupling. We show that, when the regularization is carried out in Riemann--Cartan geometry, the resulting theory admits de Sitter attractors with trace torsion on which this kinetic term remains nonvanishing. On spatially flat FLRW backgrounds with trace torsion, the quadratic actions for perturbations describe one scalar mode and two tensor modes, while the vector perturbations are nondynamical. We identify a nonempty parameter region admitting de Sitter attractors whose scalar and tensor perturbations are free of ghost and gradient instabilities at all momenta.
\end{abstract}

\maketitle
\newpage

\section{Introduction}\label{sec:intro}

The Gauss--Bonnet term gives second-order metric field equations in Riemannian geometry \cite{Lovelock} and appears as the leading curvature-squared correction to the Einstein--Hilbert action in the low-energy limit of string theory \cite{BoulwareDeser}. In dimensions higher than four, it generates nontrivial higher-curvature gravitational interactions without introducing the additional massive spin-2 ghost characteristic of generic curvature-squared theories \cite{Zwiebach}. By contrast, in four dimensions its integral is topological and its bulk variation vanishes identically \cite{Lanczos}, so with a constant coupling it does not contribute to the bulk field equations. Rescaling the coupling by $1/(D-4)$ and taking $D\to4$ was considered in Refs.~\cite{Tomozawa,Cognola} and attracted renewed interest after the proposal of Glavan and Lin \cite{GlavanLin}. Difficulties with this prescription concern the definition of a covariant four-dimensional metric theory \cite{Gurses,GursesPRL,Ai,Arrechea}, its implementation at the level of the action \cite{Mahapatra}, the assumed symmetry of the extra dimensions \cite{TianZhu}, and the singularity of its variational completion \cite{Hohmann}. Well-defined scalar--tensor formulations can instead be obtained through Kaluza--Klein reduction \cite{LuPang,Kobayashi2020,MaLu} or conformal regularization \cite{Fernandes2020,Hennigar,Easson}, the latter following the approach to the $D\to2$ limit of general relativity in Ref.~\cite{MannRoss}. These constructions are reviewed in Ref.~\cite{Fernandes2022}.

The scalar--Gauss--Bonnet theory obtained by conformal regularization is a particular shift-symmetric member of the Horndeski class \cite{Fernandes2020,Hennigar,Kobayashi2020}; see Refs.~\cite{Horndeski,DGSZ,KYY,KobayashiReview,DeffayetSteer} for the general Horndeski framework. Owing to its shift symmetry, the scalar field equation takes the form of a conservation law, and on a spatially flat Friedmann--Lema\^itre--Robertson--Walker (FLRW) background the associated current density decays as $a^{-3}$. Consequently, in an accelerating universe, the zero-current solution---which is de Sitter in vacuum---is an attractor, on which the scalar perturbation drops out of the quadratic action \cite{Kobayashi2020}. This disappearance signals strong coupling rather than decoupling of the scalar. Around flat space, which is a zero-current solution when the cosmological constant vanishes, this strong coupling has been demonstrated through tree-level scattering amplitudes \cite{BHJ}; for the zero-current cosmological solutions, it has been interpreted as a vanishing cutoff of the effective field theory \cite{Kobayashi2020,AGM1}.
The vanishing of the scalar kinetic coefficient is not specific to the Gauss--Bonnet term: for the conformally regularized Lovelock series, it persists along the zero-current solution throughout the cosmological evolution and in all physically meaningful gauges \cite{Tsujikawa2025}. The same pathology also arises outside cosmology: in the scalar--Gauss--Bonnet theory, the kinetic coefficient of one even-parity mode vanishes identically on the asymptotically flat hairy black-hole solution \cite{Tsujikawa2022}. The strong-coupling problem is therefore neither a gauge artifact nor tied to a particular background.
Nor can the problem be evaded simply by discarding the scalar while retaining four-dimensional covariance: within the class of four-dimensional Einstein--Gauss--Bonnet regularizations analyzed in Ref.~\cite{AGM1}, a theory different from general relativity must either break part of the diffeomorphism invariance or introduce additional degrees of freedom. Avoiding strong coupling in a covariant formulation therefore requires that the additional degree of freedom have a nonvanishing kinetic term on the attractor.

A geometric route to restoring the scalar dynamics is to relax the torsion-free assumption underlying the regularization. Allowing the connection to carry torsion while retaining metric compatibility leads to Riemann--Cartan geometry, the setting of Einstein--Cartan and Poincar\'e gauge gravity \cite{Hehl1976,Trautman,Shapiro,BlagojevicHehl,Obukhov,Capozziello}. Crucially, for metric-compatible connections, the four-dimensional Euler density remains topological even in the presence of torsion \cite{Nieh1980,JanssenJimenezCano2019}, so the conformal regularization can be extended without abandoning its geometric starting point. Such an extension was developed in Ref.~\cite{Qiu2026}, where the regularized Gauss--Bonnet interaction supports nontrivial torsion even in vacuum. This route need not involve additional propagating torsion modes: in the Horndeski--Cartan models studied in Refs.~\cite{MVV2023,MVV2023b,MVV2024}, the torsion perturbations are auxiliary at quadratic order about FLRW backgrounds. Nevertheless, eliminating them modifies the kinetic and gradient terms of the propagating metric and scalar modes.

Here we generalize the Riemann--Cartan regularization of Ref.~\cite{Qiu2026} to a one-parameter family of conformal transformations of the connection, all compatible with the rescaled metric. The parameter determines how the torsion trace transforms; the family includes the transformations discussed in Refs.~\cite{BuchbinderShapiro1985,Shapiro,ChakrabartyLahiri2018} and reduces to the prescription of Ref.~\cite{Qiu2026} when the contortion is left invariant. In contrast to the Horndeski--Cartan models of Refs.~\cite{MVV2023,MVV2023b,MVV2024}, whose interaction functions are free, all scalar--torsion couplings in this family are fixed by the parameter and the Gauss--Bonnet coupling. With these couplings, we find de Sitter attractors with trace torsion on which the scalar kinetic coefficient is nonzero. The reason can be read off from the conserved current, which is the product of two linear combinations of the expansion rate, the torsion trace, and the time derivative of the scalar. One of the two combinations also enters the scalar kinetic coefficient as a factor. In the torsion-free theory, the two combinations coincide. The zero-current condition then sets their common value to zero, causing the scalar kinetic coefficient to vanish as well. With torsion, the two combinations are in general distinct, so the current can vanish when one is zero while the combination entering the kinetic coefficient remains nonzero.

To determine whether this restoration is compatible with stability, we derive the scalar, vector, and tensor quadratic actions on general spatially flat FLRW backgrounds with trace torsion, using the quadratic-action methods of Refs.~\cite{Maldacena,KYY2010,KYY,DeFeliceTsujikawa2012,BelliniSawicki,DeFeliceGerardSuyama}. On these backgrounds the torsion perturbations are auxiliary, and, once all auxiliary fields are eliminated, the quadratic actions describe one propagating scalar mode and two tensor modes, while the vector sector contains no propagating degrees of freedom. In the tensor sector this elimination nevertheless makes the kinetic and gradient coefficients depend on momentum, whereas the scalar coefficients do not. For a positive coefficient of the Einstein--Cartan term, we identify a nonempty region of parameter space admitting de Sitter attractors whose scalar and tensor kinetic and gradient coefficients are positive at all momenta. The restoration of scalar dynamics is thus compatible with stability, and torsion thereby alleviates the cosmological strong coupling associated with the vanishing kinetic term of the scalar perturbation mode.

The paper is organized as follows. Section~\ref{sec:theory} presents the theory, its conventions, and the regularized action. Section~\ref{sec:background} studies FLRW backgrounds with trace torsion and identifies de Sitter solutions that act as local attractors. Section~\ref{sec:scalar} derives the quadratic action for scalar perturbations on general spatially flat FLRW backgrounds with trace torsion and compares the result on the de Sitter attractors with the torsion-free case. Sections~\ref{sec:vector} and \ref{sec:tensor} analyze the vector and tensor perturbations, respectively. Section~\ref{sec:health} identifies a region of parameter space admitting de Sitter attractors that satisfy the stability conditions. Section~\ref{sec:conclusions} discusses the results and open questions. Appendices~\ref{app:raw}--\ref{app:coupling-map} collect the expressions used in the derivations.

\section{Conformal regularization}\label{sec:theory}

We use the geometric conventions of Ref.~\cite{MVV2023} and the metric signature $(-,+,+,+)$; Greek indices $\mu,\nu,\ldots$ label spacetime components and Latin indices $i,j,k$ label spatial components. Following Ref.~\cite{Qiu2026}, a ring denotes objects built from the Levi-Civita connection of $g_{\mu\nu}$, symbols without a ring denote objects built from the full Riemann--Cartan connection, and a tilde denotes objects built from the conformally transformed metric and connection.

The full connection is metric compatible, $\nabla_\lambda g_{\mu\nu}=0$, and acts on vectors as
\begin{equation}
\nabla_\mu V^\rho=\partial_\mu V^\rho+\Gamma^\rho{}_{\mu\lambda}V^\lambda,
\qquad
\Gamma^\rho{}_{\mu\nu}=\mathring\Gamma^\rho{}_{\mu\nu}-K^\rho{}_{\mu\nu},
\label{eq:connection}
\end{equation}
where $\mathring\Gamma^\rho{}_{\mu\nu}$ is the Levi-Civita connection of $g_{\mu\nu}$ and $K$ the contortion tensor. The torsion tensor is defined by the antisymmetrized connection, and metric compatibility determines the contortion from it:
\begin{equation}
\begin{aligned}
&T^\rho{}_{\mu\nu}\equiv\Gamma^\rho{}_{\mu\nu}-\Gamma^\rho{}_{\nu\mu}=-K^\rho{}_{\mu\nu}+K^\rho{}_{\nu\mu},\\
&K^\rho{}_{\mu\nu}=-\tfrac12\bigl(T_\nu{}^\rho{}_\mu+T_\mu{}^\rho{}_\nu+T^\rho{}_{\mu\nu}\bigr),
\qquad K_{\rho\mu\nu}=-K_{\nu\mu\rho}.
\end{aligned}
\label{eq:torsion-def}
\end{equation}
The curvature of the full connection, defined as in Ref.~\cite{Trinity}, and its contractions are
\begin{equation}
R^\rho{}_{\sigma\mu\nu}=\partial_\mu\Gamma^\rho{}_{\nu\sigma}-\partial_\nu\Gamma^\rho{}_{\mu\sigma}
+\Gamma^\rho{}_{\mu\lambda}\Gamma^\lambda{}_{\nu\sigma}-\Gamma^\rho{}_{\nu\lambda}\Gamma^\lambda{}_{\mu\sigma},
\qquad
R_{\alpha\beta}\equiv R^\rho{}_{\alpha\rho\beta},
\qquad
R\equiv g^{\alpha\beta}R_{\alpha\beta},
\label{eq:curvature}
\end{equation}
so that $[\nabla_\mu,\nabla_\nu]V^\rho=R^\rho{}_{\lambda\mu\nu}V^\lambda-T^\lambda{}_{\mu\nu}\nabla_\lambda V^\rho$ and, for a scalar $\phi$, $[\nabla_\mu,\nabla_\nu]\phi=-T^\lambda{}_{\mu\nu}\partial_\lambda\phi$. With torsion, the Ricci tensor and $\nabla_\alpha\nabla_\beta\phi$ need not be symmetric, so the order of indices in the contractions below is significant.

Under the conformal rescaling $g_{\mu\nu}\to\tilde g_{\mu\nu}=e^{2\phi}g_{\mu\nu}$, the Levi-Civita connection changes by $\delta^\lambda_\beta\nabla_\alpha\phi+\delta^\lambda_\alpha\nabla_\beta\phi-g_{\alpha\beta}\nabla^\lambda\phi$, while the transformation of the contortion is not uniquely determined by the metric rescaling \cite{ChakrabartyLahiri2018}. For the conformal regularization, we construct the change of the full connection, $\Delta\Gamma^\lambda{}_{\alpha\beta}\equiv\tilde\Gamma^\lambda{}_{\alpha\beta}-\Gamma^\lambda{}_{\alpha\beta}$, in $n$ dimensions. As in the Levi-Civita connection, we take it to be linear in $\nabla_\mu\phi$, with constant coefficients and with the metric as the only other tensor. The most general parity-even form under these assumptions is
\[
\Delta\Gamma^\lambda{}_{\alpha\beta}
=A\,\delta^\lambda_\beta\nabla_\alpha\phi
+B\,\delta^\lambda_\alpha\nabla_\beta\phi
+C\,g_{\alpha\beta}\nabla^\lambda\phi,
\]
where indices are raised with $g^{\mu\nu}$. The same three-term structure is used in generalized Weyl geometry \cite{BeltranJimenezKoivisto2016,BeltranJimenezHeisenbergKoivisto2016}. All three terms are defined in any dimension $n$. In four dimensions, the totally antisymmetric tensor $\epsilon_{\lambda\alpha\beta\rho}$ permits an additional parity-odd term, $E\,\epsilon^\lambda{}_{\alpha\beta\rho}\nabla^\rho\phi$, which metric compatibility does not exclude. This term has no counterpart in $n>4$ dimensions: there $\epsilon$ carries $n$ indices, contracting it with a single gradient leaves $n-1$ of them free, and contracting any two of them with the metric gives zero. It therefore cannot enter the $n$-dimensional construction, and we do not include it. Requiring the transformed connection to remain compatible with $\tilde g_{\mu\nu}$ gives $A=1$ and $B+C=0$, leaving one free parameter, which we write as $B=1-s$. We obtain
\begin{equation}
\begin{aligned}
\tilde\Gamma^\lambda{}_{\alpha\beta}
&=\Gamma^\lambda{}_{\alpha\beta}+\delta^\lambda_\beta\nabla_\alpha\phi
+(1-s)\bigl(\delta^\lambda_\alpha\nabla_\beta\phi-g_{\alpha\beta}\nabla^\lambda\phi\bigr),\\
\tilde T^\lambda{}_{\alpha\beta}
&=T^\lambda{}_{\alpha\beta}+s\bigl(\delta^\lambda_\beta\nabla_\alpha\phi-\delta^\lambda_\alpha\nabla_\beta\phi\bigr).
\end{aligned}
\label{eq:conformal-map}
\end{equation}
Only the torsion trace $T_\alpha\equiv T^\lambda{}_{\alpha\lambda}$ changes: $\tilde T_\alpha=T_\alpha+(n-1)s\nabla_\alpha\phi$ in $n$ dimensions; in four dimensions this is the transformation of the torsion associated with the strong conformal symmetry of Refs.~\cite{BuchbinderShapiro1985,Shapiro}, with $s$ as its free parameter. The family, also discussed in Ref.~\cite{ChakrabartyLahiri2018}, includes the prescription of Ref.~\cite{Qiu2026}, in which the contortion is invariant ($s=0$), and the choice in which the connection changes by $\delta^\lambda_\beta\nabla_\alpha\phi$ alone and the torsion trace transforms as an effective Weyl gauge field ($s=1$) \cite{NiehYan1982AnnPhys,Obukhov1982}.

We define the Gauss--Bonnet scalar of the full
Riemann--Cartan connection as the Euler density,
\begin{equation}
\begin{aligned}
\mathcal G
&=
\frac14\,
\delta^{\mu_1\mu_2\mu_3\mu_4}_{\nu_1\nu_2\nu_3\nu_4}\,
R^{\nu_1\nu_2}{}_{\mu_1\mu_2}
R^{\nu_3\nu_4}{}_{\mu_3\mu_4}
\\
&=
R^{\alpha\beta\mu\nu}R_{\mu\nu\alpha\beta}
-4R_{\mu\nu}R^{\nu\mu}
+R^2 ,
\end{aligned}
\label{eq:GBdef}
\end{equation}
where
$R^{\rho\sigma}{}_{\mu\nu}
\equiv g^{\sigma\lambda}R^\rho{}_{\lambda\mu\nu}$
and
$\delta^{\mu_1\cdots\mu_4}_{\nu_1\cdots\nu_4}$
is the generalized Kronecker delta.
For a metric-compatible connection, the variation of
$\int\dd^4x\,\sqrt{-g}\,\mathcal G$ is a boundary term,
even in the presence of torsion \cite{Nieh1980,JanssenJimenezCano2019}.
With a constant coefficient this term therefore makes no
contribution to the bulk field equations.

To obtain a nontrivial four-dimensional interaction, we adopt the conformal regularization of Refs.~\cite{MannRoss,Fernandes2020,Hennigar} in its dimensional-derivative form, described in Sec.~2.3.3 of Ref.~\cite{Fernandes2022} and extended to Riemann--Cartan geometry in Ref.~\cite{Qiu2026}.
Here we apply this prescription to the one-parameter
conformal transformation \eqref{eq:conformal-map}.
We first evaluate the transformed density in $n$ dimensions,
\[
\mathcal F(n;s)
\equiv
\bigl(\sqrt{-\widetilde g}\,
\widetilde{\mathcal G}\bigr)_n ,
\]
with the same index contractions as in Eq.~\eqref{eq:GBdef}.
The dimensional derivative acts on the explicit $n$-dependence of $\mathcal F(n;s)$, with the untilded fields and $s$ held fixed. This dependence arises only from $n$-dimensional index contractions and from $\sqrt{-\widetilde g}=e^{n\phi}\sqrt{-g}$, so that $e^{-(n-4)\phi}\mathcal F(n;s)/\sqrt{-g}$ is a polynomial in $n$ and the derivative at $n=4$ exists.

The regularized action is
\begin{equation}
S
=
\int \mathrm{d}^4x
\left[
\alpha_0\sqrt{-g}
+\alpha_1\sqrt{-g}\,R
+\alpha_2
\left.
\frac{\mathrm{d}\mathcal F(n;s)}{\mathrm{d}n}
\right|_{n=4}
\right].
\label{eq:parent-action}
\end{equation}
Separating an ordinary total derivative, we write the Gauss--Bonnet term of Eq.~\eqref{eq:parent-action} as
\begin{equation}
\begin{aligned}
\mathcal L_{\mathrm{dim}}(s)
&\equiv
\left.
\frac{\mathrm{d}\mathcal F(n;s)}{\mathrm{d}n}
\right|_{n=4}
\\
&=
\sqrt{-g}\,\mathcal L_{\mathrm{rECGB}}(s)
+\partial_\mu\Theta^\mu(s),
\end{aligned}
\label{eq:dim-density}
\end{equation}
where the subscript rECGB stands for regularized Einstein--Cartan--Gauss--Bonnet and $\Theta^\mu(s)$ is a vector density. The density $\mathcal L_{\mathrm{dim}}(s)$ before integration by parts and the vector density $\Theta^\mu(s)$ are given explicitly in Eqs.~\eqref{eq:raw-split}--\eqref{eq:raw-D4} and \eqref{eq:theta} of Appendix~\ref{app:raw}.
Since $\partial_\mu\Theta^\mu(s)$ does not contribute to the field equations, all background equations and quadratic actions below are derived with the Gauss--Bonnet term of Eq.~\eqref{eq:parent-action} replaced by its bulk contribution $\sqrt{-g}\,\mathcal L_{\mathrm{rECGB}}(s)$, where
\begin{equation}
\begin{aligned}
\mathcal L_{\rm rECGB}(s)={}&
\phi\,\mathcal G
-8(s-1)G_{\alpha\beta}
 \bigl(\Phi^{\alpha\beta}-\nabla^\alpha\phi\nabla^\beta\phi\bigr)\\
&+4(s-1)^2\Big[
 (\nabla^\mu\nabla_\mu\phi)^2
 -\Phi_{\alpha\beta}\Phi^{\beta\alpha}
 +2\nabla^\alpha\phi\,\Phi_{\alpha\beta}\nabla^\beta\phi\Big]\\
&+4(s-1)^2\Big[
 XR+4sX\nabla^\mu\nabla_\mu\phi\Big]
 +8(s-1)^3(1+3s)X^2,
\end{aligned}
\label{eq:bulk-density}
\end{equation}
with
\begin{equation}
\Phi_{\alpha\beta}
\equiv\nabla_\alpha\nabla_\beta\phi,
\qquad
X\equiv-\frac12\nabla_\alpha\phi\nabla^\alpha\phi,
\qquad
G_{\alpha\beta}
\equiv R_{\alpha\beta}-\frac12g_{\alpha\beta}R .
\label{eq:bulk-defs}
\end{equation}
For $s=0$ and vanishing torsion, the regularized Gauss--Bonnet interaction reduces, up to a total derivative, to
$\alpha_2[\phi\,\mathcal G-4G^{\mu\nu}\nabla_\mu\phi\nabla_\nu\phi+8X\nabla^\mu\nabla_\mu\phi-8X^2]$. This is the scalar--Gauss--Bonnet interaction obtained by conformal regularization, with coupling $\hat\alpha=-\alpha_2$ in the convention of Ref.~\cite{Fernandes2020}; the form used in Ref.~\cite{Hennigar} is recovered after the sign flip $\phi\to-\phi$.%
\footnote{
Up to boundary terms and conventional choices of
normalization and scalar-field sign, the same scalar
interaction follows from Kaluza--Klein reduction with
a flat internal space \cite{LuPang,Kobayashi2020}.
}

\Needspace{12\baselineskip}
\section{Background cosmology}
\label{sec:background}

Spatial homogeneity and rotational isotropy allow two independent torsion
components: a trace part $T^i{}_{tj}\propto\delta^i_j$ and an axial part
$T^i{}_{jk}\propto\epsilon_{ijk}$
\cite{Tsamparlis1979,GoennerMullerHoissen1984,Iosifidis}.
We consider a spatially flat FLRW metric, a homogeneous scalar, and the trace part of torsion, where a dot denotes a derivative with respect to cosmic time $t$:
\begin{equation}
\begin{gathered}
\dd s^2=-\dd t^2+a^2(t)\delta_{ij}\dd x^i\dd x^j,
\qquad H\equiv\frac{\dot a}{a},\qquad \phi=\phi_0(t),\\
T^i{}_{tj}=u(t)\delta^i_j,\qquad T_{ijk}=0.
\end{gathered}
\label{eq:background}
\end{equation} This ansatz with trace torsion is also used in Refs.~\cite{Kranas,MVV2023}. Setting the axial torsion component to zero is consistent with its field equation, which is proportional to that component.

The metric and connection field equations on this background give the first Friedmann equation,
\begingroup
\allowdisplaybreaks[0]%
\begin{align}
0={}&\alpha_0+6\alpha_1(H^2-u^2)-6\alpha_2\ph\Big\{
4(3s-2)u^3+12su[H+(1-s)\ph]^2 \notag\\
&+6u^2[2(1-2s)H+(s-1)(3s-1)\ph]-4H^3
+6(s^2-1)H^2\ph \notag\\
&-4(s-1)^2(1+2s)H\ph^2
+(s-1)^3(1+3s)\ph^3\Big\}\equiv\mathcal C_0,
\label{eq:Friedmann}
\end{align}
\endgroup
and the torsion equation,
\begin{align}
0={}&u[\alpha_1+4\alpha_2(1-2s)H\ph
+2\alpha_2(s-1)(3s-1)\ph^2] \notag\\
&+2\alpha_2\ph\{s[H+(1-s)\ph]^2+(3s-2)u^2\}
\equiv\mathcal C_u.
\label{eq:torsion-background}
\end{align}
The scalar equation takes the form of a conservation law: the scalar appears without derivatives only in $\phi\mathcal G$, and with a constant coefficient the Euler density contributes only a boundary term (Sec.~\ref{sec:theory}), so a constant shift of $\phi$ changes the action by a boundary term. The resulting conservation law is
\begin{equation}
\frac{\dd}{\dd t}\left[a^3\alpha_2L^2\Xi\right]=0,
\label{eq:scalar-current}
\end{equation}
where
\begin{align}
L   &\equiv H-u+(1-s)\ph, \label{eq:L-eq}\\
\Xi &\equiv H+(2-3s)u+(1-s)(1+3s)\ph. \label{eq:Xi-eq}
\end{align}
On any nonstatic background ($\dot a\neq0$), the second Friedmann equation, i.e., the spatial component of the metric field equation, follows from Eqs.~\eqref{eq:Friedmann}--\eqref{eq:scalar-current} by the Noether identity associated with time reparametrizations. The background dynamics is governed by the two constraints \eqref{eq:Friedmann} and \eqref{eq:torsion-background}, which are algebraic in $(H,u,\ph)$, and the conservation law \eqref{eq:scalar-current}.

We now ask whether trace torsion can support late-time accelerated
expansion. Integrating Eq.~\eqref{eq:scalar-current} gives
\[
J(t)\equiv\alpha_2L^2\Xi=\frac{Q}{a^3(t)},
\]
where $Q$ is the conserved comoving scalar charge. Along an expanding solution with $a\to\infty$, the current density $J$ tends to zero while $Q$ remains constant, which motivates the search for late-time solutions with vanishing current. This dilution underlies the zero-current attractors found in torsion-free shift-symmetric cosmological models \cite{Kobayashi2020,KYY,DPSV2010}.

With torsion and $\alpha_2\neq0$, the zero-current condition $J=0$ requires $L=0$ or $\Xi=0$. We choose $\Xi=0$ and keep $L$ nonzero, because the kinetic coefficient of the scalar perturbation contains a factor $L^4$ (see Eq.~\eqref{eq:GSexplicit}): driving $L$ to zero would remove the kinetic term of the scalar perturbation mode, as happens in the torsion-free theory. A zero-current solution on this branch satisfies
\begin{equation}
\mathcal C_0=0,\qquad \mathcal C_u=0,\qquad \Xi=0,
\label{eq:root}
\end{equation}
which are three algebraic equations for $(H,u,\ph)$ at fixed theory parameters $(\alpha_0,\alpha_1,\alpha_2,s)$. We denote the corresponding constant values by a subscript star and retain the expanding solutions, $H_\star>0$. The background is then
\begin{equation}
a(t)=a_\star e^{H_\star t},\qquad u(t)=u_\star,\qquad
\phi_0(t)=\phi_\star+\dot\phi_{0\star}t,
\label{eq:ds-ansatz}
\end{equation}
a de Sitter metric with $\ddot a/a=H_\star^2>0$, constant trace torsion,
and a scalar that evolves linearly in time; the axial torsion vanishes.
To solve Eqs.~\eqref{eq:root}, we introduce the dimensionless ratios $p=\dot\phi_{0\star}/H_\star$ and $r=u_\star/H_\star$. For $3s\neq2$, the condition $\Xi_\star=0$ gives
\begin{equation}
r=r(p,s)
=\frac{1+(1-s)(1+3s)p}{3s-2}.
\label{eq:rmap}
\end{equation}
Substituting this into the torsion and Friedmann equations \eqref{eq:torsion-background} and \eqref{eq:Friedmann} turns them into
\begin{equation}
\frac{\alpha_1}{\alpha_2H_\star^2}=F_1(p,s),\qquad
\frac{\alpha_0}{\alpha_2H_\star^4}=F_0(p,s),
\label{eq:coupling-map}
\end{equation}
with $F_0$ and $F_1$ given in Appendix~\ref{app:coupling-map}. Defining $\beta\equiv\alpha_0\alpha_2/\alpha_1^2$ for $\alpha_1\neq0$ and eliminating $H_\star$ between the two relations \eqref{eq:coupling-map} gives
\begin{equation}
F_0(p,s)-\beta F_1(p,s)^2=0.
\label{eq:cmap}
\end{equation}
Once the couplings $(\alpha_0,\alpha_1,\alpha_2)$ and the parameter $s$ are specified, each real root $p$ of Eq.~\eqref{eq:cmap} for which $F_0(p,s)$ and $F_1(p,s)$ are well defined and $\alpha_1/[\alpha_2F_1(p,s)]>0$ yields, through Eqs.~\eqref{eq:rmap} and \eqref{eq:coupling-map}, a de Sitter solution with
\begin{equation}
H_\star=\sqrt{\frac{\alpha_1}{\alpha_2F_1(p,s)}},\qquad
u_\star=r(p,s)H_\star,\qquad \dot\phi_{0\star}=pH_\star.
\label{eq:root-components}
\end{equation}

We now establish a sufficient condition for these de Sitter solutions to be local attractors of the homogeneous background dynamics. Define the Jacobian determinant of Eqs.~\eqref{eq:root} by
\begin{equation}
\begin{aligned}
\Delta_\star
&\equiv
\det\left.\frac{\partial(\mathcal C_0,\mathcal C_u,\Xi)}
{\partial(H,u,\ph)}\right|_\star\\
&=-\frac{144H_\star\alpha_2^2\dot\phi_{0\star}(s-1)L_\star^2}{u_\star^2}
\,d_{5\star},
\end{aligned}
\label{eq:root-regularity}
\end{equation}
\begingroup
\allowdisplaybreaks[0]%
where
\begin{align}
d_{5\star}={}&s(s+1)L_\star^2(u_\star+s\dot\phi_{0\star})
+2(s-1)^2u_\star^3 \notag\\
&-2s(s-1)L_\star
\bigl(3u_\star^2+3su_\star \dot\phi_{0\star}+s^2\dot\phi_{0\star}^2\bigr).
\label{eq:d5-root}
\end{align}
\endgroup
Here $d_{5\star}$ is the value of the polynomial $d_5$ of Eq.~\eqref{eq:d5} on the de Sitter solution; the same polynomial appears in the scalar kinetic coefficient \eqref{eq:GSexplicit}. When $\Delta_\star\neq0$, the constraints \eqref{eq:Friedmann} and \eqref{eq:torsion-background} at fixed couplings define a smooth curve locally in $(H,u,\ph)$, with tangent direction $\nabla\mathcal C_0\times\nabla\mathcal C_u$. Since $\nabla\Xi\cdot(\nabla\mathcal C_0\times\nabla\mathcal C_u)=\Delta_\star\neq0$ at the solution, $\Xi$ can be used as a local coordinate on this curve, and so can $J=\alpha_2L^2\Xi$, since $\alpha_2L_\star^2\neq0$. The conservation law \eqref{eq:scalar-current} gives $\dot J=-3HJ$. Since $J_\star=0$, the linearized equation and its solution are
\begin{equation}
\delta\dot J=-3H_\star\delta J
\quad\text{and}\quad
\delta J(t)=\delta J(t_0)e^{-3H_\star(t-t_0)},
\label{eq:current-decay}
\end{equation}
where $t_0$ is the reference time at which the initial perturbation amplitude $\delta J(t_0)$ is specified.
The Lyapunov exponent $-3H_\star$ is negative. Therefore, every expanding de Sitter solution on this branch with $\Delta_\star\neq0$ is a local attractor of the homogeneous background dynamics.
\section{Scalar perturbations}\label{sec:scalar}

\subsection{Scalar perturbations on backgrounds with trace torsion}\label{sec:scalarA}

We use the scalar perturbation variables of Ref.~\cite{MVV2023}, written in cosmic time. For backgrounds with $\ph\neq0$, we impose the unitary gauge $\delta\phi=0$ and fix the spatial gauge. The metric perturbations are
\begin{equation}
\delta g_{tt}=-2\alpha,\qquad
\delta g_{ti}=a\,\partial_iB,\qquad
\delta g_{ij}=-2a^2\psi\,\delta_{ij},\qquad
\delta\phi=0,
\label{eq:metric-ansatz}
\end{equation}
and the scalar components of the contortion perturbation are
\begin{equation}
\begin{aligned}
&\delta K_{itt}=\partial_iC^{(1)},\qquad
\delta K_{itk}=\epsilon_{ikj}\partial_jC^{(5)},\\
&\delta K_{ijt}=\partial_i\partial_jC^{(2)}+\delta_{ij}C^{(3)}+\epsilon_{ijk}\partial_kC^{(4)},\\
&\delta K_{ijk}=(\delta_{ij}\partial_k-\delta_{kj}\partial_i)C^{(6)}
+\epsilon_{ikl}\partial_l\partial_jC^{(7)}
+(\epsilon_{ijl}\partial_l\partial_k-\epsilon_{kjl}\partial_l\partial_i)C^{(8)} .
\end{aligned}
\label{eq:torsion-ansatz}
\end{equation}
The metric perturbations \eqref{eq:metric-ansatz} and $C^{(1)}$, $C^{(2)}$, $C^{(3)}$, $C^{(6)}$ are even under parity, while $C^{(4)}$, $C^{(5)}$, $C^{(7)}$, $C^{(8)}$ are odd. Since the action and the background \eqref{eq:background} are invariant under spatial reflections, the two sets do not mix at quadratic order, so that $S^{\mathrm{scalar}}_2 = S_2^{\mathrm{scalar,even}} + S_2^{\mathrm{scalar,odd}}$.

Substituting Eqs.~\eqref{eq:metric-ansatz} and \eqref{eq:torsion-ansatz} into the action, expanding to second order, using the background equations, and integrating by parts, we obtain
\begingroup\small
\allowdisplaybreaks[0]
\begin{align}
S_2^{\mathrm{scalar,even}}=\frac12\int\dd t\,\dd^3x\bigg\{\,&
\frac{f_6}{a}\alpha^2-\frac32\Bigl[af_1\Ch^2+2f_2\alpha\Ch\notag\\
&\hspace{24mm}+2af_3\Ch\dot\psi+2f_4\alpha\dot\psi+af_5\dot\psi^2\Bigr] \notag\\
&+\partial_iC^{(2)}\partial_i(af_1\Ch+f_2\alpha+af_3\dot\psi) \notag\\
&+\partial_iB\,\partial_i\Big(f_3\Ch+\frac{f_4}{a}\alpha+f_5\dot\psi\Big) \notag\\
&+\partial_i\psi\,\partial_i\Big(f_7\Ch-\frac{f_5}{a}\alpha\Big)+\frac{f_8}{a}(\partial_i\psi)^2 \notag\\
&+\partial_iC^{(6)}\partial_i\left[af_9(C^{(1)}-auB)
+\frac{f_7}{a^2}\Ch-\frac{f_3}{a}\alpha+f_7\dot\psi\right]\notag\\
&+\frac{f_{10}}{a}(\partial_iC^{(6)})^2\bigg\},
\label{eq:S2}
\end{align}
\endgroup
where the ten coefficients $f_i=f_i(a,H,u,\ph,\php;s,\alpha_2)$ are listed in Eqs.~\eqref{eq:f12}--\eqref{eq:f910} of Appendix~\ref{app:coefficients}, in terms of the polynomials \eqref{eq:d0}--\eqref{eq:d5}. The explicit dependence on $\alpha_0$ and $\alpha_1$ has been eliminated using the background constraints \eqref{eq:Friedmann} and \eqref{eq:torsion-background}; this elimination requires $u\neq0$. With the same coefficients, and up to a total derivative, the parity-odd part is
\[
\begin{aligned}
S_2^{\mathrm{scalar,odd}}=\int\dd t\,\dd^3x\bigg\{&
-\frac{af_1}{4}(\partial_iC^{(4)})^2
+\frac{af_9}{2}\partial_iC^{(4)}\partial_iC^{(5)}
+\frac{f_7}{2a^2}\partial_i\partial_jC^{(4)}\partial_i\partial_jC^{(8)}\\
&-\frac{f_{10}}{a}\partial_i\partial_jC^{(7)}\partial_i\partial_jC^{(8)}
+\frac{f_{10}}{2a}(\partial_i\partial_jC^{(8)})^2\bigg\}.
\end{aligned}
\]
The parity-odd scalar action contains no time derivatives and hence describes no propagating degrees of freedom. We therefore restrict the subsequent analysis to the parity-even sector.

As in Ref.~\cite{Feng2021}, we eliminate the auxiliary fields at the level of the equations of motion: we first solve the auxiliary-field equations and then substitute the required combinations into the equation for $\psi$. Varying Eq.~\eqref{eq:S2} with respect to $C^{(1)}$, $C^{(2)}$, $C^{(3)}$, $C^{(6)}$, $B$, $\alpha$, and $\psi$, and integrating by parts, gives the following equations, in which $\nabla^2\equiv\delta^{ij}\partial_i\partial_j$ is the spatial Laplacian:
\begingroup\small
\begin{align}
C^{(1)}:&\quad f_9\lap C^{(6)}=0,
\label{eq:eom-C1}\\
C^{(2)}:&\quad \lap(af_1\Ch+f_2\alpha+af_3\dot\psi)=0,
\label{eq:eom-C2}\\
C^{(3)}:&\quad 3(af_1\Ch+f_2\alpha+af_3\dot\psi)\notag\\
&\qquad+\lap\left(af_1C^{(2)}+f_3B+f_7\psi+\frac{f_7}{a^2}C^{(6)}\right)=0,
\label{eq:eom-Q}\\
C^{(6)}:&\quad \lap\left[a^2f_9(C^{(1)}-auB)+\frac{f_7}{a}\Ch
-f_3\alpha+af_7\dot\psi+2f_{10}C^{(6)}\right]=0,
\label{eq:eom-C6}\\
B:&\quad \lap\Big(f_3\Ch+\frac{f_4}{a}\alpha+f_5\dot\psi-a^2uf_9C^{(6)}\Big)=0,
\label{eq:eom-B}\\
\alpha:&\quad 3a(f_2\Ch+f_4\dot\psi)-2f_6\alpha
+\lap(af_2C^{(2)}+f_4B-f_5\psi-f_3C^{(6)})=0,
\label{eq:eom-alpha}\\
\psi:&\quad 3\frac{\dd}{\dd t}(af_3\Ch+f_4\alpha+af_5\dot\psi)
+\frac{\dd}{\dd t}\lap(af_3C^{(2)}+f_5B+f_7C^{(6)}) \notag\\
&\quad-\frac1a\lap(af_7\Ch-f_5\alpha+2f_8\psi) \notag\\
&\quad-2a^2u\left\{3(af_1\Ch+f_2\alpha+af_3\dot\psi)
+\lap\left(af_1C^{(2)}+f_3B+f_7\psi+\frac{f_7}{a^2}C^{(6)}\right)\right\}\notag\\
&\quad=0.
\label{eq:eom-psi}
\end{align}
\endgroup
We use the comoving wavenumber $k=|\bm k|$, with physical wavenumber $k/a$, and restrict the elimination to nonzero Fourier modes. For $f_9\neq0$, which by Eq.~\eqref{eq:f910} is equivalent to $\alpha_2s\ph L\neq0$ for $u\neq0$, Eq.~\eqref{eq:eom-C1} gives $C^{(6)}=0$, and Eq.~\eqref{eq:eom-C6} then determines $C^{(1)}$. This step excludes $s=0$, which requires a separate treatment. Introduce the combinations
\[
b_1=f_1f_4-f_2f_3,\qquad
b_2=f_1f_5-f_3^2,\qquad
b_3=f_3f_4-f_2f_5 .
\]
Their explicit expressions are given in Eq.~\eqref{eq:bfactored} in Appendix~\ref{app:coefficients}. For $b_1\neq0$, equivalently $d_4\neq0$, solving Eqs.~\eqref{eq:eom-C2} and \eqref{eq:eom-B} gives
\begin{equation}
\alpha=-\frac{a\,b_2}{b_1}\dot\psi,\qquad
\Ch=-\frac{b_3}{b_1}\dot\psi.
\label{eq:alphaQ}
\end{equation}
Substituting Eq.~\eqref{eq:alphaQ} into Eqs.~\eqref{eq:eom-Q} and \eqref{eq:eom-alpha} gives
\begin{equation}
af_3\lap C^{(2)}+f_5\lap B
=\frac{b_2f_5-b_3f_7}{b_1}\lap\psi
-a\,\frac{b_2\left[3(b_1f_4-f_2b_3)+2f_6b_2\right]}{b_1^2}\dot\psi .
\label{eq:CBneeded}
\end{equation}
With Eqs.~\eqref{eq:alphaQ} and \eqref{eq:CBneeded}, Eq.~\eqref{eq:eom-psi} reduces to
\begin{equation}
G_S\ddot\psi+(3HG_S+\dot G_S)\dot\psi
-\frac{F_S}{a^2}\lap\psi=0,
\label{eq:reduced}
\end{equation}
where
\begin{equation}
G_S=\frac{b_2(3f_4b_1-3f_2b_3+2f_6b_2)}{4a^2b_1^2},
\qquad
F_S=\frac{1}{4a}\frac{\dd}{\dd t}
\left(\frac{f_5b_2-f_7b_3}{b_1}\right)-\frac{f_8}{2a^2}.
\label{eq:GSFS}
\end{equation}

The corresponding quadratic action, with its normalization inherited from Eq.~\eqref{eq:S2}, is
\begin{equation}
S_\psi=\int\dd t\,\dd^3x\,a^3
\left[G_S\dot\psi^2-\frac{F_S}{a^2}(\partial_i\psi)^2\right] .
\label{eq:cosmic-action}
\end{equation}
The no-ghost and gradient-stability conditions are
\begin{equation}
G_S>0,\qquad F_S>0 .
\label{eq:stability-scalar}
\end{equation}
Using Eqs.~\eqref{eq:f12}--\eqref{eq:bfactored}, with $d_0$, $d_4$, and $d_5$ of Eqs.~\eqref{eq:d0}, \eqref{eq:d4}, and \eqref{eq:d5}, we obtain
\begin{equation}
G_S=\frac{12\alpha_2(s-1)s^2\ph L^4}{d_4^2}\,d_5 ,
\label{eq:GSexplicit}
\end{equation}
\begin{equation}
\begin{aligned}
F_S={}&-\frac{4\alpha_2}{a}\frac{\dd}{\dd t}
\bigg\{\frac{as\ph}{u d_4}\Big[
sL^2\big(sL^2-2sLu+2(s-1)u^2\big)\\
&\hspace{34mm}
+4(s-1)u^2\big((s-1)u^2-2sLu-s^2L\ph\big)
\Big]\bigg\}
+\frac{4\alpha_2}{u}\big(\ph d_0-2u\php\big).
\end{aligned}
\label{eq:FSexplicit}
\end{equation}

Evaluating Eqs.~\eqref{eq:GSexplicit} and \eqref{eq:FSexplicit} on the de Sitter branch $\Xi_\star=0$ gives
\begin{equation}
\begin{aligned}
G_{S\star}&=\frac{12\alpha_2(s-1)s^2\dot\phi_{0\star} L_\star^4}{d_{4\star}^2}\,d_{5\star},\\[4pt]
F_{S\star}&=\frac{4\alpha_2\dot\phi_{0\star}L_\star}{9d_{4\star}}
\Big[s(9s+4)L_\star^3-6s(s-2)L_\star^2u_\star\\
&\qquad-18s(s-1)(s-2)L_\star u_\star^2
-36(s-1)^2u_\star^3\Big].
\end{aligned}
\label{eq:GSroot}
\end{equation}
At $s=1$, $\Xi$ and $L$ coincide, so $\Xi_\star=0$ implies $L_\star=0$ and this case lies outside the branch considered here. 
Combining Eqs.~\eqref{eq:root-regularity} and \eqref{eq:GSroot} yields
\begin{equation}
G_{S\star}=-\frac{\Delta_\star s^2L_\star^2u_\star^2}{12H_\star\alpha_2d_{4\star}^2}.
\label{eq:GSDelta}
\end{equation}
Equation~\eqref{eq:GSDelta} shows that the condition $\Delta_\star\neq0$, which makes the de Sitter solution a local attractor (Sec.~\ref{sec:background}), also guarantees a nonzero scalar kinetic coefficient. For $H_\star>0$, the no-ghost condition $G_{S\star}>0$ is equivalent to $\alpha_2\Delta_\star<0$. Section~\ref{sec:health} identifies a nonempty parameter region in which each point admits an expanding de Sitter solution with a positive scalar kinetic coefficient.

\Needspace{10\baselineskip}
\subsection{Scalar perturbations in the torsion-free theory}\label{sec:torsionfree}

For comparison, we consider the torsion-free theory with the regularized scalar--Gauss--Bonnet interaction \cite{LuPang,Kobayashi2020,Fernandes2020,Hennigar}, obtained by setting $s=0$ and the contortion to zero in the action.
We work in vacuum and show directly that the scalar kinetic and gradient coefficients vanish on the de Sitter attractor. After integration by parts, the action \eqref{eq:parent-action} becomes
\begin{equation}
\begin{aligned}
S^{(s=0)}_{\text{torsion-free}}=\int\dd^4x\sqrt{-g}\,\Big\{\alpha_0+\alpha_1\mathring R
+\alpha_2\Big[&\phi\mathring{\mathcal G}-4\mathring G^{\mu\nu}\mathring\nabla_\mu\phi\mathring\nabla_\nu\phi\\
&-4(\mathring\nabla\phi)^2\mathring\Box\phi-2\big((\mathring\nabla\phi)^2\big)^2\Big]\Big\}.
\end{aligned}
\label{eq:tf-action}
\end{equation}
Here the ring denotes quantities constructed from the Levi-Civita connection, as in Sec.~\ref{sec:theory}, and $\mathring\Box\phi\equiv g^{\mu\nu}\mathring\nabla_\mu\mathring\nabla_\nu\phi$. On a spatially flat FLRW background, $L=\Xi=H+\ph$, and the background equations reduce to
\begin{equation}
\alpha_0+6\alpha_1H^2-6\alpha_2H^4+6\alpha_2L^4=0,
\qquad
\frac{\dd}{\dd t}\big(a^3\alpha_2L^3\big)=0.
\label{eq:tf-background}
\end{equation}

We first derive the scalar quadratic action on a general background satisfying Eq.~\eqref{eq:tf-background}. We use the metric perturbations and unitary gauge of Eq.~\eqref{eq:metric-ansatz}, with $\ph\neq0$. The lapse and shift perturbations $\alpha$ and $B$ are auxiliary fields. Expanding Eq.~\eqref{eq:tf-action} to second order and using the background equations gives
\begin{equation}
\begin{aligned}
S_2^{(s=0),\mathrm{scalar}}=\frac12\int\dd t\,\dd^3x\,\bigg\{&
\frac{\fz6}{a}\alpha^2-\frac32\big[2\fz4\alpha\dot\psi+a\fz5\dot\psi^2\big]\\
&+\partial_iB\,\partial_i\bigg(\frac{\fz4}{a}\alpha+\fz5\dot\psi\bigg)
-\frac{\fz5}{a}\partial_i\psi\,\partial_i\alpha
+\frac{\fz8}{a}(\partial_i\psi)^2\bigg\},
\end{aligned}
\label{eq:tf-S2}
\end{equation}
up to boundary terms. 
Without contortion perturbations, the scalar action has no parity-odd part.
We use the same coefficient notation as in Eq.~\eqref{eq:S2}; the superscript $(0)$ denotes coefficients calculated directly in the torsion-free theory: 
\begin{equation}
\begin{aligned}
\fz4 &= 8a^3\big[H(\alpha_1-2\alpha_2H^2)+2\alpha_2L^3\big],\\
\fz5 &= 8a^2\big[\alpha_1-2\alpha_2H^2+2\alpha_2L^2\big],\\
\fz6 &= -12a^4\big[H^2(\alpha_1-2\alpha_2H^2)+2\alpha_2L^4\big],\\
\fz8 &= 4a^2\big[\alpha_1-2\alpha_2\ph^2+4\alpha_2\ddot\phi_0\big].
\end{aligned}
\label{eq:tf-coeffs}
\end{equation}

Varying Eq.~\eqref{eq:tf-S2} with respect to $B$ and $\alpha$ gives the constraints
\begin{equation}
\begin{aligned}
\nabla^2\bigg(\frac{\fz4}{a}\alpha+\fz5\dot\psi\bigg)&=0,\\
3a\fz4\dot\psi-2\fz6\alpha
+\nabla^2\big(\fz4B-\fz5\psi\big)&=0.
\end{aligned}
\label{eq:tf-constraints}
\end{equation}
For a Fourier mode with $k\neq0$ and $\fz4\neq0$, they determine
\begin{equation}
\alpha=-\frac{a\fz5}{\fz4}\dot\psi,
\qquad
\nabla^2 B=-a\left(3+\frac{2\fz6\fz5}{(\fz4)^2}\right)\dot\psi
+\frac{\fz5}{\fz4}\nabla^2\psi.
\label{eq:tf-alphaB}
\end{equation}
Substitution into Eq.~\eqref{eq:tf-S2}, followed by integration by parts in time, gives the reduced action in the form of Eq.~\eqref{eq:cosmic-action}, with
\begin{equation}
G_S^{(0)}=\frac{\fz5\big[3(\fz4)^2+2\fz5\fz6\big]}{4a^2(\fz4)^2},
\qquad
F_S^{(0)}=\frac1{4a}\frac{\dd}{\dd t}\left(\frac{(\fz5)^2}{\fz4}\right)
-\frac{\fz8}{2a^2}.
\label{eq:tf-reduced}
\end{equation}

On the torsion-free de Sitter attractor, $L_\star=0$ and $\dot\phi_{0\star}=-H_\star$. Evaluating Eq.~\eqref{eq:tf-reduced} on this background gives
\begin{equation}
G_{S\star}^{(0)}=F_{S\star}^{(0)}=0.
\label{eq:tf-zero}
\end{equation}
Thus the reduced scalar quadratic action vanishes on the attractor, in agreement with Ref.~\cite{Kobayashi2020}.

\section{Vector perturbations}\label{sec:vector}

Following the decomposition of Ref.~\cite{MVV2023}, the vector sector contains the two metric perturbations $S_i$ and $F_i$ and the six contortion perturbations $\Vn1_i,\dots,\Vn6_i$. 
In cosmic time, the perturbations are
\begin{equation}
\delta g_{tt}=0,\qquad
\delta g_{ti}=aS_i,\qquad
\delta g_{ij}=a^2(\partial_iF_j+\partial_jF_i),
\label{eq:vmetric}
\end{equation}
\begin{equation}
\begin{aligned}
\delta K_{itt}&=\Vn1_i,\qquad
\delta K_{ijt}=\partial_i\Vn2_j+\partial_j\Vn3_i,\qquad
\delta K_{itk}=\partial_i\Vn4_k-\partial_k\Vn4_i,\\
\delta K_{ijk}&=\delta_{ij}\Vn5_k-\delta_{kj}\Vn5_i
+\partial_j\partial_i\Vn6_k-\partial_j\partial_k\Vn6_i.
\end{aligned}
\label{eq:vtorsion}
\end{equation}
All vector variables are transverse. Under transverse spatial diffeomorphisms $x^i\to x^i+\xi^i$, with $\partial_i\xi_i=0$, the vector variables transform as
\begin{equation}
\begin{aligned}
&S_i\to S_i-a\dot\xi_i,\qquad F_i\to F_i-\xi_i,\\
&\Vn1_i\to\Vn1_i-a^2u\dot\xi_i,\\
&\Vn2_i\to\Vn2_i-a^2u\xi_i,\qquad
\Vn3_i\to\Vn3_i-a^2u\xi_i,
\end{aligned}
\label{eq:vgauge}
\end{equation}
while $\Vn4_i$, $\Vn5_i$, and $\Vn6_i$ are invariant.
The seven local combinations
\begin{equation}
\begin{aligned}
&\hS_i\equiv S_i-a\dot F_i,\qquad
\hV1_i\equiv\Vn1_i-auS_i,\\
&\hV2_i\equiv\Vn2_i-a^2uF_i,\qquad
\hV3_i\equiv\Vn3_i-a^2uF_i,\\
&\hV4_i\equiv\Vn4_i,\qquad
\hV5_i\equiv\Vn5_i,\qquad
\hV6_i\equiv\Vn6_i
\end{aligned}
\label{eq:vinvariants}
\end{equation}
are invariant. Together with $F_i$, they are related to the original variables by a local invertible transformation,\footnote{The inverse is $S_i=\hS_i+a\dot F_i$, $\Vn1_i=\hV1_i+au(\hS_i+a\dot F_i)$, $\Vn2_i=\hV2_i+a^2uF_i$, and $\Vn3_i=\hV3_i+a^2uF_i$, with $\Vn4_i=\hV4_i$, $\Vn5_i=\hV5_i$, and $\Vn6_i=\hV6_i$.} so that solutions of the Euler--Lagrange equations in the two sets of variables are in one-to-one correspondence \cite{Takahashi2017}. Expanding the action to second order, using the background equations and integrating by parts, we obtain
\begin{equation}
\begin{aligned}
S_2^{V}=\int\dd t\,\dd^3x\bigg\{&
\frac{f_5}{16a}(\partial_j\hS_i)^2
+\frac{f_3}{8}\partial_j\hS_i\,\partial_j(\hV2_i+\hV3_i)
+\frac{f_7}{8a}\partial_j\hS_i\partial_j\hV5_i\\
&+\frac{af_9}{2}\hV1_i\hV5_i
-\frac{af_9}{4}\partial_j\hV1_i\partial_j\hV6_i
+\frac{af_1}{4}\partial_j\hV2_i\partial_j\hV3_i\\
&+\frac{af_9}{4}\partial_j(\hV2_i-\hV3_i)\partial_j\hV4_i
+\frac{f_7}{4a^2}\partial_j\hV2_i\partial_j\hV5_i\\
&+\frac{f_{10}}{2a}\hV5_i\hV5_i
-\frac{f_{10}}{2a}\partial_j\hV5_i\partial_j\hV6_i
\bigg\},
\end{aligned}
\label{eq:SV}
\end{equation}
with the same $f_1,f_3,f_5,f_7,f_9,f_{10}$ as in the scalar sector. The action \eqref{eq:SV} depends only on the invariants \eqref{eq:vinvariants} and contains no time derivatives of these invariants. The vector sector therefore carries no propagating degrees of freedom on FLRW backgrounds with trace torsion at quadratic order.

\Needspace{9\baselineskip}
\section{Tensor perturbations}\label{sec:tensor}

The tensor sector contains the metric perturbation $h_{ij}$ and two tensor components of the contortion perturbation, $\TOne_{ij}$ and $\TTwo_{ij}$. Using the decomposition of Ref.~\cite{MVV2023} in cosmic time, we write
\begin{equation}
\begin{aligned}
&\delta g_{tt}=0,\qquad \delta g_{ti}=0,\qquad
\delta g_{ij}=2a^2h_{ij},\\
&\delta K_{ijt}=\TOne_{ij},\qquad
\delta K_{ijk}=\partial_i\TTwo_{jk}-\partial_k\TTwo_{ji}.
\end{aligned}
\label{eq:tensor-ansatz}
\end{equation}
The tensors $h_{ij}$, $\TOne_{ij}$, and $\TTwo_{ij}$ are symmetric, transverse, and traceless. We introduce the combination
\begin{equation}
U_{ij}\equiv\TOne_{ij}-2a^2u\,h_{ij}.
\label{eq:Udef}
\end{equation}
Expanding the action to second order, using the background equations and integrating by parts, we obtain the quadratic action for tensor perturbations:
\begin{equation}
\begin{aligned}
S_2^{\mathrm{tensor}}=\frac18\int\dd t\,\dd^3x\bigg\{&
af_5\,\dot h_{ij}\dot h_{ij}-\frac{2f_8}{a}\,\partial_kh_{ij}\partial_kh_{ij}
-2af_3\,\dot h_{ij}U_{ij}+af_1\,U_{ij}U_{ij}\\
&+2f_7\,\partial_kh_{ij}\partial_kU_{ij}
+\frac{2f_7}{a^2}\,\partial_kU_{ij}\,\partial_k\TTwo_{ij}
-\frac{2f_{10}}{a}\,\partial_k\TTwo_{ij}\partial_k\TTwo_{ij}\\
&-2f_7\,\partial_k\dot h_{ij}\,\partial_k\TTwo_{ij}\bigg\},
\end{aligned}
\label{eq:ST}
\end{equation}
where the coefficients $f_i$ are the same as in the scalar sector. Neither $U_{ij}$ nor $\TTwo_{ij}$ carries a time derivative.

Varying Eq.~\eqref{eq:ST} with respect to $U_{ij}$, $\TTwo_{ij}$, and $h_{ij}$ gives
\begin{align}
U_{ij}:&\quad af_1U_{ij}-af_3\,\dot h_{ij}
-f_7\lap\left(h_{ij}+\frac{\TTwo_{ij}}{a^2}\right)=0,
\label{eq:eom-U}\\
\TTwo_{ij}:&\quad \lap\left[\frac{2f_{10}}{a}\TTwo_{ij}+f_7\,\dot h_{ij}
-\frac{f_7}{a^2}U_{ij}\right]=0,
\label{eq:eom-T2}\\
h_{ij}:&\quad \frac{\dd}{\dd t}\left[af_5\,\dot h_{ij}-af_3U_{ij}
+f_7\lap\TTwo_{ij}\right]
-\frac{2f_8}{a}\lap h_{ij}+f_7\lap U_{ij}=0 .
\label{eq:eom-h}
\end{align}
In Fourier space, where $\lap\to-k^2$, Eqs.~\eqref{eq:eom-U} and \eqref{eq:eom-T2} form an algebraic system for $U_{ij}$ and $\TTwo_{ij}$:
\begin{equation}
\begin{pmatrix}
af_1 & \dfrac{f_7}{a^2}k^2\\[8pt]
\dfrac{f_7}{a^2}k^2 & -\dfrac{2f_{10}}{a}k^2
\end{pmatrix}
\begin{pmatrix}U_{ij}\\[2pt]\TTwo_{ij}\end{pmatrix}
=\begin{pmatrix}af_3\,\dot h_{ij}-f_7k^2h_{ij}\\[2pt]f_7k^2\dot h_{ij}\end{pmatrix},
\qquad
\det=-k^2D_T,
\label{eq:matrix}
\end{equation}
\begin{equation}
D_T\equiv 2f_1f_{10}+\frac{f_7^{\,2}}{a^4}k^2 .
\label{eq:DT}
\end{equation}
For $k\neq0$ and $D_T\neq0$, solving this system and using Eq.~\eqref{eq:Udef} and the identity $a^2f_1-f_3=a^2f_9$ of Eq.~\eqref{eq:identities} gives
\begin{equation}
\begin{aligned}
\TOne_{ij}&=2a^2u\,h_{ij}+a^2\dot h_{ij}
-\frac{2f_{10}}{aD_T}\Bigl[a^3f_9\,\dot h_{ij}+f_7k^2h_{ij}\Bigr],
\\
\TTwo_{ij}&=-\frac{f_7}{a^2D_T}\Bigl[a^3f_9\,\dot h_{ij}+f_7k^2h_{ij}\Bigr].
\end{aligned}
\label{eq:T12}
\end{equation}
Thus both contortion perturbations are determined by $h_{ij}$ and $\dot h_{ij}$.

Substituting Eq.~\eqref{eq:T12} into Eq.~\eqref{eq:eom-h} and using the coefficient identities in Appendix~\ref{app:coefficients}, we obtain
\begin{equation}
\begin{gathered}
\frac{\dd}{\dd t}\left(a^3G_T\dot h_{ij}\right)
+aF_Tk^2h_{ij}=0,\\
G_T=-\frac{a^2f_9^{\,2}f_{10}}{4D_T}=\frac{f_{10}\,b_2}{4a^2D_T},
\qquad
F_T=\frac{e_1+e_2k^2}{D_T^{\,2}},
\end{gathered}
\label{eq:GTFT}
\end{equation}
with
\begin{equation}
\begin{aligned}
e_1&=\frac{f_{10}^2}{2a^2}\left[2f_1^2f_8
+af_1\frac{\dd(f_7f_3)}{\dd t}-af_7f_3\dot f_1\right],\\
e_2&=-\frac{f_1f_{10}^{\,2}f_7^{\,2}}{2a^2}
-\frac{f_7^{\,2}}{4a^9}\left[f_7\frac{\dd(a^6f_{10}f_9)}{\dd t}
-a^6f_{10}f_9\dot f_7\right].
\end{aligned}
\label{eq:e0e2}
\end{equation}
Substituting Eq.~\eqref{eq:T12} into Eq.~\eqref{eq:ST} and integrating by parts, we obtain the reduced quadratic action for tensor perturbations:
\begin{equation}
S_2^{\mathrm{tensor}}=\int\dd t\,\intk a^3\bigg[
G_T\,\dot h_{ij}(t,\bm k)\dot h_{ij}(t,-\bm k)
-\frac{F_T}{a^2}k^2\,h_{ij}(t,\bm k)h_{ij}(t,-\bm k)\bigg].
\label{eq:red-action}
\end{equation}
Expanding $h_{ij}$ in the polarization tensors of Ref.~\cite{Feng2021},
\begin{equation}
h_{ij}(t,\bm k)=\sum_{\lambda=+,\times}e^\lambda_{ij}(\bm k)\,h_\lambda(t,\bm k),
\qquad
k_ie^\lambda_{ij}=e^\lambda_{ii}=0,\qquad
e^\lambda_{ij}(\bm k)\,e^{\rho}_{ij}(\bm k)^*=2\delta_{\lambda\rho},
\label{eq:polarization}
\end{equation}
with $e^\lambda_{ij}(-\bm k)=e^\lambda_{ij}(\bm k)^*$ and $h_\lambda(t,-\bm k)=h_\lambda(t,\bm k)^*$, we obtain
\begin{equation}
S_2^{\mathrm{tensor}}=2\sum_{\lambda=+,\times}\int\dd t\,\intk a^3\bigg[
G_T\,|\dot h_\lambda(t,\bm k)|^2
-\frac{F_T}{a^2}k^2\,|h_\lambda(t,\bm k)|^2\bigg],
\label{eq:red-action-pol}
\end{equation}
and each polarization obeys
\begin{equation}
\ddot h_\lambda+\left(3H+\frac{\dot G_T}{G_T}\right)\dot h_\lambda+c_T^2\frac{k^2}{a^2}h_\lambda=0,
\qquad
c_T^2=\frac{F_T}{G_T}.
\label{eq:polarization-eom}
\end{equation}
For $G_T\neq0$, the reduced action describes two propagating tensor modes with the same kinetic and gradient coefficients. These coefficients depend on momentum through the elimination of the contortion perturbations: $U_{ij}$ enters Eq.~\eqref{eq:ST} both without derivatives and through gradient couplings to $h_{ij}$ and $\TTwo_{ij}$, so the factor $D_T$ in the determinant of the auxiliary system \eqref{eq:matrix} depends on $k$. The same mechanism produces momentum-dependent tensor kinetic and gradient coefficients in the $G_5$ sector of the Horndeski--Cartan theory studied in Ref.~\cite{MVV2024}.

\section{Stability of de Sitter solutions}
\label{sec:health}

Stability of the expanding de Sitter solutions of Sec.~\ref{sec:background} requires positive kinetic and gradient coefficients for the scalar and tensor modes \cite{KYY,MVV2024}; the vector sector has no propagating mode at quadratic order. With $q=k/a$ the physical momentum, these conditions read
\begin{equation}
G_S>0,\qquad F_S>0,\qquad
G_T(q)>0,\qquad F_T(q)>0
\qquad\text{for all }q>0.
\label{eq:requirement}
\end{equation}
Equation~\eqref{eq:GTFT} can be written as
\begin{equation}
G_T(q)=\frac{A_T}{1+x},\qquad
F_T(q)=\frac{B_T+C_Tx}{(1+x)^2},\qquad
x=\frac{q^2}{q_\star^2},
\label{eq:GTFTx}
\end{equation}
with
\begin{equation}
\begin{aligned}
q_\star^2&=\frac{2a^2f_1f_{10}}{f_7^{\,2}},\qquad
&A_T&=-\frac{a^2f_9^{\,2}}{8f_1},\\
B_T&=\frac{e_1}{(2f_1f_{10})^2},\qquad
&C_T&=\frac{a^2q_\star^2e_2}{(2f_1f_{10})^2}.
\end{aligned}
\label{eq:ABC}
\end{equation}
The tensor conditions in Eq.~\eqref{eq:requirement} therefore hold if
\begin{equation}
q_\star^2>0,\qquad A_T>0,\qquad B_T>0,\qquad C_T>0.
\label{eq:ABCpos}
\end{equation}

\begin{figure}[!htbp]
\centering
\includegraphics[width=0.58\textwidth]{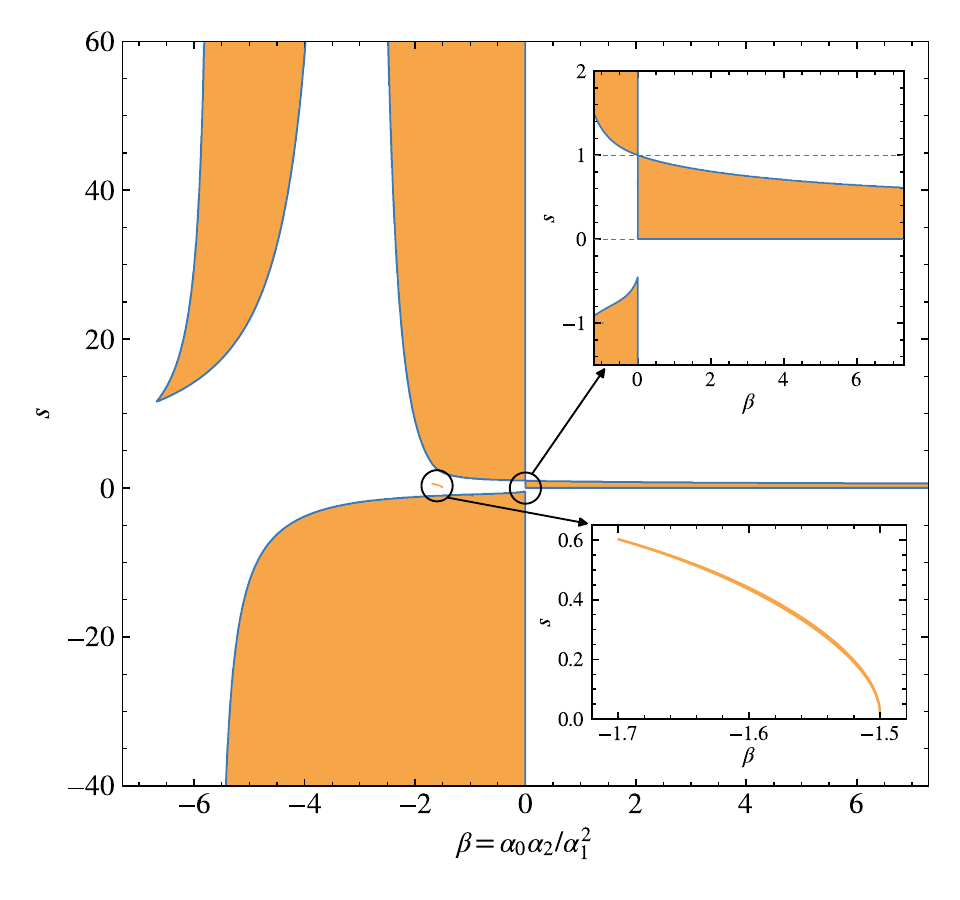}
\caption{Region in the $(\beta,s)$ plane admitting at least one expanding de Sitter attractor satisfying Eq.~\eqref{eq:requirement}.}
\label{fig:region}
\end{figure}

For each $(\beta,s)$, we identify the real roots of Eq.~\eqref{eq:cmap} that give expanding de Sitter solutions satisfying Eq.~\eqref{eq:requirement}, with $G_S$ and $F_S$ given by Eq.~\eqref{eq:GSroot}. Since $G_S>0$ implies $\Delta_\star\neq0$ by Eq.~\eqref{eq:GSDelta}, each such solution is a local attractor of the homogeneous background dynamics (Sec.~\ref{sec:background}). Figure~\ref{fig:region} shows where at least one such solution exists.

\FloatBarrier

\Needspace{8\baselineskip}
\section{Conclusions and discussion}\label{sec:conclusions}

We have applied the conformal regularization of Refs.~\cite{Fernandes2020,Hennigar,Fernandes2022} to the Gauss--Bonnet term in Riemann--Cartan geometry, extending our earlier work \cite{Qiu2026} to a one-parameter family of conformal transformations of the connection. Within the class considered in Sec.~\ref{sec:theory}, this family is the most general one compatible with the rescaled metric, and $s=0$ reproduces the prescription of Ref.~\cite{Qiu2026}. Our main result is that, with trace torsion, the theory admits de Sitter attractors on which the scalar kinetic coefficient remains nonzero. The attractor property itself is not specific to torsion: the expanding de Sitter solutions constructed in Sec.~\ref{sec:background} are local attractors of the homogeneous background dynamics whenever $\Delta_\star\neq0$, because the shift-symmetry current density $J=\alpha_2L^2\Xi$ decays as $a^{-3}$, as in the torsion-free shift-symmetric models of Refs.~\cite{KYY,DPSV2010}. The difference from the torsion-free case lies in how this current density vanishes on the attractor. With torsion, $L$ and $\Xi$ are distinct, so $J$ can vanish through $\Xi_\star=0$ while $L_\star$ remains nonzero, and Eq.~\eqref{eq:GSDelta} shows that $\Delta_\star\neq0$ then guarantees a nonzero scalar kinetic coefficient. In the torsion-free theory, $L$ and $\Xi$ coincide, so the vanishing current density sets the scalar kinetic coefficient to zero on the attractor and the scalar perturbation drops out of the quadratic action \cite{Kobayashi2020}, signaling strong coupling.

To determine whether these attractors are stable, we have derived the quadratic actions for scalar, vector, and tensor perturbations about general spatially flat FLRW backgrounds with trace torsion. After the constraints are solved and the auxiliary variables are eliminated, one scalar mode and two tensor modes propagate, while the vector sector carries no propagating degrees of freedom. Torsion therefore adds no propagating mode at this order, but eliminating the torsion perturbations makes the tensor kinetic and gradient coefficients depend on momentum. Applying these actions to the de Sitter attractors, we have identified, for $\alpha_1>0$, a nonempty region of parameter space that admits de Sitter attractors whose scalar and tensor kinetic and gradient coefficients are positive at all momenta.

Our perturbative results, obtained at quadratic order on FLRW backgrounds, open several directions for further study. A first question is whether the absence of propagating torsion modes persists in the full nonlinear theory. Establishing the nonlinear degree-of-freedom count requires a Hamiltonian analysis of the action, following the methods of Refs.~\cite{LangloisNoui2016,BlagojevicNikolic1983,YoNester1999}. A second question concerns strong coupling. A nonvanishing scalar kinetic coefficient does not by itself exclude strong coupling; determining the scale at which the interactions of the canonically normalized scalar mode become strong requires the cubic and higher-order actions and is left for future work.

Beyond the de Sitter attractors, the background and perturbation equations derived here can be used to determine which initial conditions evolve toward these attractors and whether the perturbations remain stable along the way. Extending the analysis to include matter and radiation would allow us to determine whether the attractors can be preceded by radiation- and matter-dominated eras, as in Ref.~\cite{MartinMoruno2015b}. Finally, one may ask whether the theory admits nonsingular, geodesically complete histories that end on one of these attractors. Nonsingular cosmologies have been constructed in Horndeski--Cartan models by choosing the interaction functions to produce a prescribed evolution \cite{MVV2023b,MVV2024}. In the present theory these functions are fixed by the regularization, so the question requires a separate analysis.

\Needspace{7\baselineskip}
\begin{acknowledgments}
This work was supported in part by the National Natural Science Foundation of China (NSFC) under Grant No.~12547104.
\end{acknowledgments}

\appendix

\section{The regularized Lagrangian density before integration by parts and the resulting boundary term}\label{app:raw}
Evaluating the dimensional derivative in Eq.~\eqref{eq:parent-action} for the conformal transformation \eqref{eq:conformal-map} gives the Lagrangian density \eqref{eq:dim-density} term by term, without integration by parts and with all torsion terms written explicitly,
\begin{equation}
\mathcal L_{\rm dim}(s)=\sqrt{-g}\,
\bigl[\mathcal L^{(s^0)}_{\rm rGB}+\mathcal L^{(s^1)}_{\rm rGB}+\mathcal L^{(s^2)}_{\rm rGB}+\mathcal L^{(s^3)}_{\rm rGB}+\mathcal L^{(s^4)}_{\rm rGB}\bigr].
\label{eq:raw-split}
\end{equation}
Here the subscript rGB denotes the density before integration by parts. Indices are raised and lowered with $g^{\mu\nu}$ and $g_{\mu\nu}$, the index order of the curvature is that of Eq.~\eqref{eq:curvature}, the torsion is defined in Eq.~\eqref{eq:torsion-def}, and $\nabla_\alpha\nabla_\beta\phi\equiv\nabla_\alpha(\nabla_\beta\phi)$, as in $\Phi_{\alpha\beta}$ of Eq.~\eqref{eq:bulk-defs}. The first three terms of $\mathcal L^{(s^0)}_{\rm rGB}$ combine to  $\phi\,\mathcal G$, and each $\mathcal L^{(s^k)}_{\rm rGB}$ carries an explicit factor $s^k$:
\allowdisplaybreaks
\begin{align}
\mathcal L^{(s^0)}_{\rm rGB}={}& + \phi\,R^{\alpha\beta\gamma\delta}\,R_{\gamma\delta\alpha\beta} - 4\,\phi\,R^{\alpha\beta}{}_{\alpha}{}^{\gamma}\,R_{\gamma}{}^{\delta}{}_{\beta\delta} + \phi\,R^{\alpha\beta}{}_{\alpha\beta}\,R^{\gamma\delta}{}_{\gamma\delta} - 4\,R^{\beta\gamma}{}_{\beta\gamma}\,\nabla_{\alpha}\nabla^{\alpha}\phi \nonumber\\
& - 4\,\phi\,R^{\beta\gamma}{}_{\beta\gamma}\,\nabla_{\alpha}\nabla^{\alpha}\phi - 8\,\phi\,T_{\alpha}{}^{\beta\gamma}\,R_{\beta}{}^{\delta}{}_{\gamma\delta}\,\nabla^{\alpha}\phi - 4\,\phi\,T^{\beta\gamma\delta}\,R_{\gamma\delta\alpha\beta}\,\nabla^{\alpha}\phi \nonumber\\
& - 8\,\phi\,T^{\beta}{}_{\beta}{}^{\gamma}\,R_{\gamma}{}^{\delta}{}_{\alpha\delta}\,\nabla^{\alpha}\phi + 8\,\phi\,T^{\beta}{}_{\alpha}{}^{\gamma}\,R_{\gamma}{}^{\delta}{}_{\beta\delta}\,\nabla^{\alpha}\phi - 4\,\phi\,T^{\beta}{}_{\alpha\beta}\,R^{\gamma\delta}{}_{\gamma\delta}\,\nabla^{\alpha}\phi \nonumber\\
& - 2\,R^{\beta\gamma}{}_{\beta\gamma}\,\nabla_{\alpha}\phi\,\nabla^{\alpha}\phi - 8\,T^{\gamma}{}_{\beta\gamma}\,\nabla_{\alpha}\nabla^{\beta}\phi\,\nabla^{\alpha}\phi + 12\,\nabla_{\alpha}\nabla^{\alpha}\phi\,\nabla_{\beta}\nabla^{\beta}\phi \nonumber\\
& + 8\,\phi\,\nabla_{\alpha}\nabla^{\alpha}\phi\,\nabla_{\beta}\nabla^{\beta}\phi + 8\,T^{\gamma}{}_{\alpha\gamma}\,\nabla^{\alpha}\phi\,\nabla_{\beta}\nabla^{\beta}\phi + 8\,\phi\,T^{\gamma}{}_{\alpha\gamma}\,\nabla^{\alpha}\phi\,\nabla_{\beta}\nabla^{\beta}\phi \nonumber\\
& + 20\,\nabla_{\alpha}\phi\,\nabla^{\alpha}\phi\,\nabla_{\beta}\nabla^{\beta}\phi + 8\,\phi\,\nabla_{\alpha}\phi\,\nabla^{\alpha}\phi\,\nabla_{\beta}\nabla^{\beta}\phi + 4\,\phi\,T_{\alpha}{}^{\gamma\delta}\,T_{\beta\gamma\delta}\,\nabla^{\alpha}\phi\,\nabla^{\beta}\phi \nonumber\\
& + 8\,\phi\,T_{\alpha}{}^{\gamma\delta}\,T_{\gamma\beta\delta}\,\nabla^{\alpha}\phi\,\nabla^{\beta}\phi + 8\,\phi\,T_{\alpha\beta}{}^{\gamma}\,T^{\delta}{}_{\gamma\delta}\,\nabla^{\alpha}\phi\,\nabla^{\beta}\phi - 8\,R_{\alpha}{}^{\gamma}{}_{\beta\gamma}\,\nabla^{\alpha}\phi\,\nabla^{\beta}\phi \nonumber\\
& - 8\,\phi\,R_{\alpha}{}^{\gamma}{}_{\beta\gamma}\,\nabla^{\alpha}\phi\,\nabla^{\beta}\phi + 12\,T^{\gamma}{}_{\beta\gamma}\,\nabla_{\alpha}\phi\,\nabla^{\alpha}\phi\,\nabla^{\beta}\phi + 8\,\phi\,T^{\gamma}{}_{\beta\gamma}\,\nabla_{\alpha}\phi\,\nabla^{\alpha}\phi\,\nabla^{\beta}\phi \nonumber\\
& + 6\,\nabla_{\alpha}\phi\,\nabla^{\alpha}\phi\,\nabla_{\beta}\phi\,\nabla^{\beta}\phi + 24\,\nabla^{\alpha}\phi\,\nabla_{\beta}\nabla_{\alpha}\phi\,\nabla^{\beta}\phi + 16\,\phi\,\nabla^{\alpha}\phi\,\nabla_{\beta}\nabla_{\alpha}\phi\,\nabla^{\beta}\phi \nonumber\\
& - 8\,\phi\,T^{\gamma}{}_{\beta\gamma}\,\nabla^{\alpha}\phi\,\nabla^{\beta}\nabla_{\alpha}\phi + 8\,\phi\,R_{\alpha}{}^{\gamma}{}_{\beta\gamma}\,\nabla^{\beta}\nabla^{\alpha}\phi + 8\,R_{\beta}{}^{\gamma}{}_{\alpha\gamma}\,\nabla^{\beta}\nabla^{\alpha}\phi \nonumber\\
& - 4\,\nabla_{\alpha}\nabla_{\beta}\phi\,\nabla^{\beta}\nabla^{\alpha}\phi - 8\,\nabla_{\beta}\nabla_{\alpha}\phi\,\nabla^{\beta}\nabla^{\alpha}\phi - 8\,\phi\,\nabla_{\beta}\nabla_{\alpha}\phi\,\nabla^{\beta}\nabla^{\alpha}\phi \nonumber\\
& + 8\,T_{\alpha\beta\gamma}\,\nabla^{\alpha}\phi\,\nabla^{\gamma}\nabla^{\beta}\phi - 8\,\phi\,T_{\beta\alpha\gamma}\,\nabla^{\alpha}\phi\,\nabla^{\gamma}\nabla^{\beta}\phi - 8\,T_{\gamma\alpha\beta}\,\nabla^{\alpha}\phi\,\nabla^{\gamma}\nabla^{\beta}\phi, \label{eq:raw-D0}
\end{align}
\begin{align}
\mathcal L^{(s^1)}_{\rm rGB}={}& s\,\Bigl[\, + 4\,R^{\beta\gamma}{}_{\beta\gamma}\,\nabla_{\alpha}\nabla^{\alpha}\phi + 4\,\phi\,R^{\beta\gamma}{}_{\beta\gamma}\,\nabla_{\alpha}\nabla^{\alpha}\phi + 4\,\phi\,T^{\beta\gamma\delta}\,R_{\gamma\delta\alpha\beta}\,\nabla^{\alpha}\phi \nonumber\\
& + 8\,\phi\,T^{\beta}{}_{\beta}{}^{\gamma}\,R_{\gamma}{}^{\delta}{}_{\alpha\delta}\,\nabla^{\alpha}\phi - 8\,\phi\,T^{\beta}{}_{\alpha}{}^{\gamma}\,R_{\gamma}{}^{\delta}{}_{\beta\delta}\,\nabla^{\alpha}\phi + 4\,\phi\,T^{\beta}{}_{\alpha\beta}\,R^{\gamma\delta}{}_{\gamma\delta}\,\nabla^{\alpha}\phi \nonumber\\
& + 4\,R^{\beta\gamma}{}_{\beta\gamma}\,\nabla_{\alpha}\phi\,\nabla^{\alpha}\phi + 16\,T^{\gamma}{}_{\beta\gamma}\,\nabla_{\alpha}\nabla^{\beta}\phi\,\nabla^{\alpha}\phi + 8\,\phi\,T^{\gamma}{}_{\beta\gamma}\,\nabla_{\alpha}\nabla^{\beta}\phi\,\nabla^{\alpha}\phi \nonumber\\
& - 24\,\nabla_{\alpha}\nabla^{\alpha}\phi\,\nabla_{\beta}\nabla^{\beta}\phi - 16\,\phi\,\nabla_{\alpha}\nabla^{\alpha}\phi\,\nabla_{\beta}\nabla^{\beta}\phi - 16\,T^{\gamma}{}_{\alpha\gamma}\,\nabla^{\alpha}\phi\,\nabla_{\beta}\nabla^{\beta}\phi \nonumber\\
& - 16\,\phi\,T^{\gamma}{}_{\alpha\gamma}\,\nabla^{\alpha}\phi\,\nabla_{\beta}\nabla^{\beta}\phi - 60\,\nabla_{\alpha}\phi\,\nabla^{\alpha}\phi\,\nabla_{\beta}\nabla^{\beta}\phi - 24\,\phi\,\nabla_{\alpha}\phi\,\nabla^{\alpha}\phi\,\nabla_{\beta}\nabla^{\beta}\phi \nonumber\\
& - 8\,\phi\,T_{\alpha}{}^{\gamma\delta}\,T_{\gamma\beta\delta}\,\nabla^{\alpha}\phi\,\nabla^{\beta}\phi - 8\,\phi\,T_{\alpha\beta}{}^{\gamma}\,T^{\delta}{}_{\gamma\delta}\,\nabla^{\alpha}\phi\,\nabla^{\beta}\phi + 16\,R_{\alpha}{}^{\gamma}{}_{\beta\gamma}\,\nabla^{\alpha}\phi\,\nabla^{\beta}\phi \nonumber\\
& + 16\,\phi\,R_{\alpha}{}^{\gamma}{}_{\beta\gamma}\,\nabla^{\alpha}\phi\,\nabla^{\beta}\phi - 36\,T^{\gamma}{}_{\beta\gamma}\,\nabla_{\alpha}\phi\,\nabla^{\alpha}\phi\,\nabla^{\beta}\phi - 24\,\phi\,T^{\gamma}{}_{\beta\gamma}\,\nabla_{\alpha}\phi\,\nabla^{\alpha}\phi\,\nabla^{\beta}\phi \nonumber\\
& - 24\,\nabla_{\alpha}\phi\,\nabla^{\alpha}\phi\,\nabla_{\beta}\phi\,\nabla^{\beta}\phi - 72\,\nabla^{\alpha}\phi\,\nabla_{\beta}\nabla_{\alpha}\phi\,\nabla^{\beta}\phi - 48\,\phi\,\nabla^{\alpha}\phi\,\nabla_{\beta}\nabla_{\alpha}\phi\,\nabla^{\beta}\phi \nonumber\\
& + 8\,\phi\,T^{\gamma}{}_{\beta\gamma}\,\nabla^{\alpha}\phi\,\nabla^{\beta}\nabla_{\alpha}\phi - 8\,R_{\beta}{}^{\gamma}{}_{\alpha\gamma}\,\nabla^{\beta}\nabla^{\alpha}\phi - 8\,\phi\,R_{\beta}{}^{\gamma}{}_{\alpha\gamma}\,\nabla^{\beta}\nabla^{\alpha}\phi \nonumber\\
& + 16\,\nabla_{\alpha}\nabla_{\beta}\phi\,\nabla^{\beta}\nabla^{\alpha}\phi + 8\,\nabla_{\beta}\nabla_{\alpha}\phi\,\nabla^{\beta}\nabla^{\alpha}\phi + 16\,\phi\,\nabla_{\beta}\nabla_{\alpha}\phi\,\nabla^{\beta}\nabla^{\alpha}\phi \nonumber\\
& - 8\,T_{\alpha\beta\gamma}\,\nabla^{\alpha}\phi\,\nabla^{\gamma}\nabla^{\beta}\phi - 16\,\phi\,T_{\alpha\beta\gamma}\,\nabla^{\alpha}\phi\,\nabla^{\gamma}\nabla^{\beta}\phi + 8\,\phi\,T_{\beta\alpha\gamma}\,\nabla^{\alpha}\phi\,\nabla^{\gamma}\nabla^{\beta}\phi \nonumber\\
& + 16\,T_{\gamma\alpha\beta}\,\nabla^{\alpha}\phi\,\nabla^{\gamma}\nabla^{\beta}\phi + 8\,\phi\,T_{\gamma\alpha\beta}\,\nabla^{\alpha}\phi\,\nabla^{\gamma}\nabla^{\beta}\phi\,\Bigr], \label{eq:raw-D1}
\end{align}
\begin{align}
\mathcal L^{(s^2)}_{\rm rGB}={}& s^2\,\Bigl[\, - 2\,R^{\beta\gamma}{}_{\beta\gamma}\,\nabla_{\alpha}\phi\,\nabla^{\alpha}\phi - 8\,T^{\gamma}{}_{\beta\gamma}\,\nabla_{\alpha}\nabla^{\beta}\phi\,\nabla^{\alpha}\phi - 8\,\phi\,T^{\gamma}{}_{\beta\gamma}\,\nabla_{\alpha}\nabla^{\beta}\phi\,\nabla^{\alpha}\phi \nonumber\\*
& + 12\,\nabla_{\alpha}\nabla^{\alpha}\phi\,\nabla_{\beta}\nabla^{\beta}\phi + 8\,\phi\,\nabla_{\alpha}\nabla^{\alpha}\phi\,\nabla_{\beta}\nabla^{\beta}\phi + 8\,T^{\gamma}{}_{\alpha\gamma}\,\nabla^{\alpha}\phi\,\nabla_{\beta}\nabla^{\beta}\phi \nonumber\\
& + 8\,\phi\,T^{\gamma}{}_{\alpha\gamma}\,\nabla^{\alpha}\phi\,\nabla_{\beta}\nabla^{\beta}\phi + 60\,\nabla_{\alpha}\phi\,\nabla^{\alpha}\phi\,\nabla_{\beta}\nabla^{\beta}\phi + 24\,\phi\,\nabla_{\alpha}\phi\,\nabla^{\alpha}\phi\,\nabla_{\beta}\nabla^{\beta}\phi \nonumber\\
& - 8\,R_{\alpha}{}^{\gamma}{}_{\beta\gamma}\,\nabla^{\alpha}\phi\,\nabla^{\beta}\phi - 8\,\phi\,R_{\alpha}{}^{\gamma}{}_{\beta\gamma}\,\nabla^{\alpha}\phi\,\nabla^{\beta}\phi + 36\,T^{\gamma}{}_{\beta\gamma}\,\nabla_{\alpha}\phi\,\nabla^{\alpha}\phi\,\nabla^{\beta}\phi \nonumber\\
& + 24\,\phi\,T^{\gamma}{}_{\beta\gamma}\,\nabla_{\alpha}\phi\,\nabla^{\alpha}\phi\,\nabla^{\beta}\phi + 36\,\nabla_{\alpha}\phi\,\nabla^{\alpha}\phi\,\nabla_{\beta}\phi\,\nabla^{\beta}\phi \nonumber\\
& + 72\,\nabla^{\alpha}\phi\,\nabla_{\beta}\nabla_{\alpha}\phi\,\nabla^{\beta}\phi + 48\,\phi\,\nabla^{\alpha}\phi\,\nabla_{\beta}\nabla_{\alpha}\phi\,\nabla^{\beta}\phi - 12\,\nabla_{\alpha}\nabla_{\beta}\phi\,\nabla^{\beta}\nabla^{\alpha}\phi \nonumber\\
& - 8\,\phi\,\nabla_{\alpha}\nabla_{\beta}\phi\,\nabla^{\beta}\nabla^{\alpha}\phi - 8\,T_{\gamma\alpha\beta}\,\nabla^{\alpha}\phi\,\nabla^{\gamma}\nabla^{\beta}\phi - 8\,\phi\,T_{\gamma\alpha\beta}\,\nabla^{\alpha}\phi\,\nabla^{\gamma}\nabla^{\beta}\phi\,\Bigr], \label{eq:raw-D2}
\end{align}
\begin{align}
\mathcal L^{(s^3)}_{\rm rGB}={}& s^3\,\Bigl[\, - 20\,\nabla_{\alpha}\phi\,\nabla^{\alpha}\phi\,\nabla_{\beta}\nabla^{\beta}\phi - 8\,\phi\,\nabla_{\alpha}\phi\,\nabla^{\alpha}\phi\,\nabla_{\beta}\nabla^{\beta}\phi - 12\,T^{\gamma}{}_{\beta\gamma}\,\nabla_{\alpha}\phi\,\nabla^{\alpha}\phi\,\nabla^{\beta}\phi \nonumber\\
& - 8\,\phi\,T^{\gamma}{}_{\beta\gamma}\,\nabla_{\alpha}\phi\,\nabla^{\alpha}\phi\,\nabla^{\beta}\phi - 24\,\nabla_{\alpha}\phi\,\nabla^{\alpha}\phi\,\nabla_{\beta}\phi\,\nabla^{\beta}\phi \nonumber\\
& - 24\,\nabla^{\alpha}\phi\,\nabla_{\beta}\nabla_{\alpha}\phi\,\nabla^{\beta}\phi - 16\,\phi\,\nabla^{\alpha}\phi\,\nabla_{\beta}\nabla_{\alpha}\phi\,\nabla^{\beta}\phi\,\Bigr], \label{eq:raw-D3}
\end{align}
\begin{align}
\mathcal L^{(s^4)}_{\rm rGB}={}& s^4\,\Bigl[\, + 6\,\nabla_{\alpha}\phi\,\nabla^{\alpha}\phi\,\nabla_{\beta}\phi\,\nabla^{\beta}\phi\,\Bigr]. \label{eq:raw-D4}
\end{align}

After integration by parts, Eqs.~\eqref{eq:raw-split}--\eqref{eq:raw-D4} take the form
\begin{equation}
\mathcal L_{\rm dim}(s)=\sqrt{-g}\,\mathcal L_{\rm rECGB}(s)+\partial_\mu\Theta^\mu(s),
\label{eq:theta-def}
\end{equation}
\begin{equation}
\begin{aligned}
\Theta^\mu(s)=\sqrt{-g}\Bigl[&8(s-1)^2(1+\phi)\bigl(\nabla^\mu\phi\,\nabla_\alpha\nabla^\alpha\phi-\nabla_\nu\phi\,\Phi^{\nu\mu}\bigr)\\
&+8(s-1)^3(3+2\phi)\,X\nabla^\mu\phi-8(s-1)\,\phi\,G^{\mu\nu}\nabla_\nu\phi\Bigr],
\end{aligned}
\label{eq:theta}
\end{equation}
which is the boundary term of Eq.~\eqref{eq:dim-density}. 
\section{Coefficients of the quadratic actions}\label{app:coefficients}

The coefficients below are written on the branch $u\neq0$, where Eq.~\eqref{eq:torsion-background} determines $\alpha_1$. The polynomials $d_i$ organize the background dependence, the coefficients $f_i$ enter all three perturbation sectors, and the combinations $b_i$ control the scalar auxiliary-field equations. 

\subsection{Polynomials \texorpdfstring{$d_i$}{d\_i}}\label{app:di-background}
With $L$ defined in Eq.~\eqref{eq:L-eq}, we introduce the following polynomials in the background quantities:
\begin{align}
d_0={}&sL^2-2(s-1)Lu-2su^2-2s(s-1)u\ph,
\label{eq:d0}\\
d_1={}&-sL^2-2sLu+2(s-1)u^2,
\label{eq:d1}\\
d_2={}&2sL^3+2L^2[s(s-1)\ph-(s+1)u]
+8(s-1)Lu^2+4(s-1)^2u^2\ph,
\label{eq:d2}\\
d_3={}&6sL^4+12L^3[s(s-1)\ph-u] \notag\\
&+6L^2[8(s-1)u^2+(s-1)(3s-1)u\ph+s(s-1)^2\ph^2] \notag\\
&+36(s-1)^2Lu^2\ph+12(s-1)^3u^2\ph^2,
\label{eq:d3}\\
d_4={}&s^2L^3+s(s-1)L^2(u+s\ph)
+2s(s-1)Lu(2u+s\ph)-2(s-1)^2u^3,
\label{eq:d4}\\
d_5={}&s(s+1)L^2(u+s\ph)
-2s(s-1)L(3u^2+3su\ph+s^2\ph^2)+2(s-1)^2u^3.
\label{eq:d5}
\end{align}

\subsection{Coefficients \texorpdfstring{$f_i$}{f\_i} and combinations \texorpdfstring{$b_i$}{b\_i}}\label{app:fi-and-bi-perturbation}
The ten coefficients $f_1,\dots,f_{10}$ of the quadratic actions \eqref{eq:S2}, \eqref{eq:SV} and \eqref{eq:ST} are
\begingroup\small
\begin{align}
f_1&=-\frac{16\alpha_2\ph}{a^2u}d_1,
&f_2&=32a\alpha_2\ph[sL^2-2(s-1)Lu-(s-1)^2u\ph],
\label{eq:f12}\\
f_3&=32\alpha_2\ph[sL-(s-1)u],
&f_4&=-\frac{8a^3\alpha_2\ph}{u}d_2,
\label{eq:f34}\\
f_5&=-\frac{16a^2\alpha_2\ph}{u}[sL^2-2sLu+2(s-1)u^2],
&f_6&=\frac{4a^4\alpha_2\ph}{u}d_3,
\label{eq:f56}\\
f_7&=-\frac{32\alpha_2s\ph}{a},
&f_8&=\frac{8a^2\alpha_2}{u}(2u\php-\ph d_0),
\label{eq:f78}\\
f_9&=\frac{16\alpha_2s\ph L^2}{a^2u},
&f_{10}&=\frac{8\alpha_2}{a^2u}
\{2(s-1)u\php 
+\ph L(sL+2u)\},
\label{eq:f910}
\end{align}
\endgroup
and the three combinations of Sec.~\ref{sec:scalar} are
\begin{equation}
\begin{aligned}
&b_1=-\frac{256\alpha_2^2\ph^2L^2a}{u^2}d_4,\qquad
b_2=-\frac{256\alpha_2^2s^2\ph^2}{u^2}L^4,\\
&b_3=\frac{512\alpha_2^2(s-1)\ph^2L^2a^3}{u}
[(s-1)u^2-2sLu-s^2L\ph].
\end{aligned}
\label{eq:bfactored}
\end{equation}

The coefficients in Eqs.~\eqref{eq:f12}--\eqref{eq:f910} satisfy
\begin{equation}
\begin{aligned}
f_5&=2a^2f_3-a^4f_1,\\
2f_8+2a^4f_{10}+a\frac{\dd}{\dd t}\bigl(a^2f_7\bigr)&=0,\\
a^2f_1-f_3&=a^2f_9.
\end{aligned}
\label{eq:identities}
\end{equation}
The first and third identities give
\begin{equation}
b_2=f_1f_5-f_3^2=-(a^2f_1-f_3)^2=-a^4f_9^2.
\label{eq:b2identity}
\end{equation}

\section{Functions relating the couplings to the de Sitter solutions}\label{app:coupling-map}
The functions $F_1$ and $F_0$ of Eq.~\eqref{eq:coupling-map} are
\begin{align}
F_1(p,s) &=
\frac{2p(s-1)^2\bigl[3+(5+6s)p+(2+4s+3s^2)p^2\bigr]}
{-1-(1+2s-3s^2)p},
\label{eq:F1} \\[0.3em]
F_0(p,s)&=\frac{-6p(s-1)^3\,\mathcal Q(p,s)}{(3s-2)^2\,(-1-p-2ps+3ps^2)},
\label{eq:F0}
\end{align}
where
\begin{align}
\mathcal Q(p,s)={}&-18-36p-24p^2-8p^3-2p^4+54s+36ps-48p^2s-52p^3s-14p^4s \notag\\
&+108ps^2+18p^2s^2-75p^3s^2-29p^4s^2
+162p^2s^3+63p^3s^3-9p^4s^3 \notag\\
&+108p^3s^4+27p^4s^4+27p^4s^5 .
\label{eq:Qpoly}
\end{align}

\renewcommand{\bibfont}{\small\linespread{1.45}\selectfont}
\renewcommand{\bibsection}{\section*{\refname}}
\bibliography{references}

\end{document}